\documentclass[english,aps,pre,preprint,superscriptaddress]{revtex4}
\usepackage[T1]{fontenc}
\usepackage[utf8]{inputenc}
\usepackage{amsmath}
\usepackage{graphicx}
\usepackage{amssymb}
\begin{document}

\title{Maier--Saupe model for a mixture of uniaxial and biaxial molecules}
\author{E. S. Nascimento}
\affiliation{Instituto de F\'{\i}sica, Universidade de S\~{a}o Paulo, 
Caixa Postal 66318, 05314-970, S\~{a}o Paulo, SP, Brazil}
\author{E. F. Henriques}
\affiliation{Instituto de F\'{\i}sica e Matem\'{a}tica, 
Universidade Federal de Pelotas, Caixa Postal 354, 96010-900, 
Pelotas, RS, Brasil}
\author{A. P. Vieira}
\affiliation{Instituto de F\'{\i}sica, Universidade de S\~{a}o Paulo, 
Caixa Postal 66318, 05314-970, S\~{a}o Paulo, SP, Brazil}
\author{S. R. Salinas}
\affiliation{Instituto de F\'{\i}sica, Universidade de S\~{a}o Paulo, 
Caixa Postal 66318, 05314-970, S\~{a}o Paulo, SP, Brazil}

\begin{abstract}
We introduce shape fluctuations in a liquid-crystalline system by considering
an elementary Maier--Saupe lattice model for a mixture of uniaxial and biaxial
molecules. Shape variables are treated in the annealed (thermalized) limit. We
analyze the thermodynamic properties of this system in terms of temperature
$T$, concentration $c$ of intrinsically biaxial molecules, and a parameter
$\Delta$ associated with the degree of biaxiality of the molecules. At the
mean-field level, we use standard techniques of statistical mechanics to draw
global phase diagrams, which are shown to display a rich structure, including
uniaxial and biaxial nematic phases, a reentrant ordered region, and many
distinct multicritical points. Also, we use the formalism to write an
expansion of the free energy in order to make contact with the Landau--de
Gennes theory of nematic phase transitions.

\end{abstract}
\maketitle

\section{Introduction}

The characterization of biaxial nematic phases in a number of thermotropic
liquid-crystalline systems \cite{Acharya, Madsen, Merkel}
stimulated a revival of interest in the investigation of theoretical models to
describe biaxial structures \cite{LuckhurstNature}. About forty years ago,
Freiser \cite{Freiser} showed the existence of uniaxial and biaxial nematic
phases in a generalization of the mean-field Maier--Saupe theory of the
nematic transition with the addition of suitably asymmetric degrees of
freedom. A nematic biaxial phase has also been shown to exist in a lattice
model with steric interactions between platelets \cite{Alben}, and in a number
of calculations for model systems with soft and hard-core interactions
\cite{Bocarra1970, Straley, Luckhurst2012}. The early experimental
results, however, referred to a lyotropic liquid-crystalline mixture
\cite{YuSaupe}, which should be better represented by a model of uniaxial
nematogenic elements \cite{BerardiZannoni, MartinezRaton}, and which
motivated the use of an elementary version of the Maier--Saupe theory
\cite{Henriques, ECarmo, ECarmo2011} to investigate a lattice
statistical model for a binary mixture of cylinders and disks. We now propose
an extension of this elementary model, along the lines of Freiser's
generalization of the Maier--Saupe theory, in order to analyze the global
phase diagram of a mixture of uniaxial and biaxial molecules.

In some analytical \cite{LongaPajakWydro} and numerical \cite{BerardiZannoni}
calculations, it has been pointed out that shape fluctuations play an
important role in the stability of the biaxial nematic phases. In spite of the
complexity of the liquid-crystalline systems, whose complete description may
require the introduction of more realistic, and necessarily involved,
theoretical models, we believe that there is still room for the investigation
of elementary statistical lattice models, with the addition of some
ingredients that may be essential to describe the main features of the
thermodynamic behavior. Along the lines of Freiser's early work, we then add
extra degrees of freedom, of biaxial nature, to an elementary lattice model,
which leads to the definition of a six-state Maier--Saupe (MS6) model. This
MS6 model is similar to an earlier proposal by Bocarra and collaborators
\cite{Bocarra1970}, and may be regarded as a generalization of a previously
used three-state Potts model to describe the uniaxial nematic transition
\cite{OliveiraFigueiredo}. Shape fluctuations are taken into account by
introducing a \textquotedblleft biaxiality parameter\textquotedblright%
\ $\Delta$, and by considering a binary mixture of molecules with $\Delta=0$
(intrinsically uniaxial molecules) and $\Delta\neq0$ (intrinsically biaxial
molecules). This model system is sufficiently simple to be amenable to
detailed statistical mechanics calculations for either quenched \cite{Ma} or
annealed mixtures of molecules. We then carry out calculations to obtain
global phase diagrams, and write an expansion of the free energy to make
contact with the standard form of the phenomenological Landau--de Gennes
theory of phase transitions.

This paper is divided as follows. In Section \ref{sec:MS6}, we define the MS6
model for a binary mixture of molecules, and formulate the statistical
problem. In Section \ref{sec:mean-field}, we analyze the mean-field equations,
draw a number of characteristic phase diagrams, and make contact with the
Landau--de Gennes theory. Section \ref{sec:conc} is devoted to a final
discussion and to some conclusions.

\section{The six-state Maier--Saupe model}

\label{sec:MS6}

The standard formulations of the Maier--Saupe theory of nematic phase
transitions \cite{deGennes} can be described in terms of the Hamiltonian
\begin{equation}
\mathcal{H}=-\epsilon\sum_{\left(  i,j\right)  }\sum_{\alpha,\beta
=1,2,3}\Omega_{i}^{\alpha\beta}\,\Omega_{j}^{\alpha\beta}, \label{energia_par}%
\end{equation}
where $\epsilon$ is a positive parameter, $\left(  i,j\right)  $ means that
the sum is over pairs of molecules at sites $i$ and $j$, and $\mathbf{\Omega
}_{i}$ is the symmetric traceless quadrupole tensor associated with a molecule
at site $i$. In general, $\mathbf{\Omega}_{i}$ may be written in terms of the
direction cosines or Euler angles which connect the laboratory and molecular
frames. From the traceless condition, we write the eigenvalues $\Lambda
_{1}=-1+\Delta$, $\Lambda_{2}=-1-\Delta$, and $\Lambda_{3}=2$, of the tensor
$\Omega$, where $\Delta$ is a parameter that gauges the degree of biaxiality.

The problem is considerably simplified if we resort to a discretization of
directions, which has been used to describe the isotropic-nematic transition
\cite{OliveiraFigueiredo}. We then assume that the principal molecular axes
are restricted to the directions of the Cartesian coordinates of the
laboratory. Therefore, the quadrupole tensor $\mathbf{\Omega}$ can assume only
six states represented by the matrices
\begin{equation}
\left(
\begin{array}
[c]{ccc}%
-1+\Delta & 0 & 0\\
0 & -1-\Delta & 0\\
0 & 0 & 2
\end{array}
\right)  ,\left(
\begin{array}
[c]{ccc}%
-1+\Delta & 0 & 0\\
0 & 2 & 0\\
0 & 0 & -1-\Delta
\end{array}
\right)  , \nonumber\label{seisa}%
\end{equation}%
\begin{equation}
\left(
\begin{array}
[c]{ccc}%
-1-\Delta & 0 & 0\\
0 & -1+\Delta & 0\\
0 & 0 & 2
\end{array}
\right)  ,\left(
\begin{array}
[c]{ccc}%
-1-\Delta & 0 & 0\\
0 & 2 & 0\\
0 & 0 & -1+\Delta
\end{array}
\right)  , \label{seisb}%
\end{equation}%
\begin{equation}
\left(
\begin{array}
[c]{ccc}%
2 & 0 & 0\\
0 & -1-\Delta & 0\\
0 & 0 & -1+\Delta
\end{array}
\right)  ,\left(
\begin{array}
[c]{ccc}%
2 & 0 & 0\\
0 & -1+\Delta & 0\\
0 & 0 & -1-\Delta
\end{array}
\right)  , \nonumber\label{seisc}%
\end{equation}
which leads to the definition of the six-state Maier--Saupe (MS6) model. If
the molecules are intrinsically uniaxial ($\Delta=0$), we regain a three-state
model, which has been used to describe the transition from the isotropic to
the uniaxial nematic phase \cite{OliveiraFigueiredo}, and to investigate the
existence of a biaxial nematic phase in a binary mixture of cylinders and
disks \cite{Henriques, ECarmo, ECarmo2011}.

The thermodynamic behavior of the MS6 model is determined from the canonical
partition function
\begin{equation}
Z=\sum_{\left\{  \mathbf{\Omega}_{i}\right\}  }\exp\left[  \beta\epsilon
\sum_{\left(  i,j\right)  }\sum_{\alpha=1,2,3}\Omega_{i}^{\alpha\alpha
}\,\Omega_{j}^{\alpha\alpha}\right]  ,
\end{equation}
where $\beta\epsilon=1/T$, so that $T$ is the temperature in suitable units,
and the first sum is over all microscopic configurations $\left\{
\mathbf{\Omega}_{i}\right\}  $ of this six-state model. This problem is
further simplified if we consider a fully connected model, with equal
interactions between all pairs of sites. At this mean-field level, if
$\Delta=0$, we anticipate just a first-order transition between an isotropic
and a uniaxial nematic phase. If $\Delta\neq0$, however, we can describe the
transition to a stable biaxial nematic phase.

We now turn to a mixture of intrinsically uniaxial ($\Delta=0$) and
intrinsically biaxial ($\Delta\neq0$) molecules. In this mixture we have two
sets of degrees of freedom: (i) orientational degrees of freedom, $\left\{
\mathbf{\Omega}_{i}\right\}  $, of quadrupolar nature, and (ii)
shape-disordered degrees of freedom, $\left\{  \Delta_{i}\right\}  $, with
either $\Delta_{i}=0$ or $\Delta_{i}=\Delta\neq0$, at all lattice sites. These
two sets of degrees of freedom may be associated with quite different
relaxation times, which leads to the distinction between annealed and quenched
situations \cite{Ma, Witten}. In the quenched case, the
\textquotedblleft shape-disordered\textquotedblright\ degrees of freedom never
reach thermal equilibrium during the experimental times. Given a configuration
$\left\{  \Delta_{i}\right\}  $, we calculate a partition function $Z=Z\left(
\left\{  \Delta_{i}\right\}  \right)  $, and a configuration-dependent free
energy, $f\left(  \left\{  \Delta_{i}\right\}  \right)  $. The free energy of
the system is an average of $f\left(  \left\{  \Delta_{i}\right\}  \right)  $
over the shape-disordered degrees of freedom, and the concentration of
intrinsically biaxial molecules is not a true variable of equilibrium
thermodynamics. In the annealed case, the two sets of degrees of freedom are
supposed to thermalize during the experimental time, so that concentration and
chemical potential are thermodynamically conjugate variables. In the annealed
case, given the concentration, both types of particles are free to move across
the system in order to minimize the free energy. In this work, we consider
annealed disorder only, which is more appropriate to a liquid-crystalline system.

In the annealed case, consider a binary mixture of $N_{1}$ intrinsically
biaxial molecules ($\Delta\neq0$) and $N_{2}=N-N_{1}$ uniaxial molecules
($\Delta=0$). Given the numbers of uniaxial and biaxial molecules, the
canonical partition function is a sum over orientational and disorder
configurations,
\begin{equation}
Z_{a}=\underset{\left\{  \mathbf{\Omega}_{i}\right\}  }{\sum}\underset
{\left\{  \Delta_{i}\right\}  }{\sum}^{\prime}\exp\left[  \beta\epsilon
\sum_{\left(  i,j\right)  }\sum_{\alpha=1,2,3}\Omega_{i}^{\alpha\alpha}\left(
\Delta_{i}\right)  \,\Omega_{j}^{\alpha\alpha}\left(  \Delta_{j}\right)
\right]  ,
\end{equation}
where the prime in the second sum indicates the restriction%
\begin{equation}
\sum_{i=1}^{N}\Delta_{i}=N_{1}\Delta\quad.
\end{equation}
At this stage, it is convenient to introduce a chemical potential $\mu$ and
change to a grand ensemble. First, we redefine the shape variable of molecule
$i$ such that
\begin{equation}
\Delta_{i}=n_{i}\Delta,
\end{equation}
where
\begin{equation}
n_{i}=%
\begin{cases}
0, & \text{for a uniaxial object,}\\
1, & \text{for a biaxial object.}%
\end{cases}
\end{equation}
Then, the grand partition function is given by
\begin{equation}
\Xi_{a}=\underset{\left\{  \mathbf{\Omega}_{i}\right\}  }{\sum}\underset
{\left\{  n_{i}\right\}  }{\sum}\exp\left[  \beta\epsilon\sum_{\left(
i,j\right)  }\sum_{\alpha=1,2,3}\Omega_{i}^{\alpha\alpha}\left(  n_{i}\right)
\,\Omega_{j}^{\alpha\alpha}\left(  n_{j}\right)  +\beta\mu\sum_{i}%
n_{i}\right]  , \label{grandcanonical}%
\end{equation}
where $\mu$ is the chemical potential that controls the number of biaxial
molecules. We now remark that the sums over configurations in
Eq.(\ref{grandcanonical}) are no longer restricted, which makes it possible to
carry out the calculations in the mean-field limit, as it is detailed in the
next Section.

\section{Mean-field calculations}

\label{sec:mean-field}

The mean-field version of the MS6 model is given by the Hamiltonian
\begin{equation}
\label{h_campo_medio}%
\begin{split}
\mathcal{H}_{MF}  &  =-\frac{\epsilon}{2N}\sum_{i,j=1}^{N}\sum_{\alpha=1}%
^{3}\Omega_{i}^{\alpha\alpha}\left(  n_{i}\right)  \, \Omega_{j}^{\alpha
\alpha}\left(  n_{j}\right) \\
&  =-\frac{\epsilon}{2N}\sum_{\alpha=1}^{3}\left[  \sum_{i=1}^{N}\Omega
_{i}^{\alpha\alpha} \left(  n_{i}\right)  \right]  ^{2}.
\end{split}
\end{equation}

The grand partition function $\Xi_{a}$ in Eq. (\ref{grandcanonical}) can be
factorized by using three Gaussian identities,
\begin{equation}
\exp\left[  \frac{\beta\epsilon}{2N}\left(  \sum_{i=1}^{N}\Omega_{i}%
^{\alpha\alpha}(n_{i})\right)  ^{2}\right]  \propto\int_{-\infty}^{+\infty
}dq_{\alpha}\exp\left[  -\frac{\beta\epsilon N}{2}q_{\alpha}^{2}+\beta\epsilon
q_{\alpha}\sum_{i=1}^{N}\Omega_{i}^{\alpha\alpha}(n_{i})\right]  ,
\end{equation}
with $\alpha\in\{1,2,3\}$. This factorization effectively decouples the
problem of calculating the grand partition function, and the sums over
$\{\mathbf{\Omega}_{i}\}$ and $\{n_{i}\}$ can be performed in a
straightforward way, so that we can write
\begin{equation}
\Xi_{a}\propto\int_{-\infty}^{+\infty}dq_{1}\int_{-\infty}^{+\infty}dq_{2}%
\int_{-\infty}^{+\infty}dq_{3}\exp(-N\beta\epsilon\psi), \label{grandpot}%
\end{equation}
where $\psi$ a functional of $\left\{  q_{\alpha}\right\}  $. In the
thermodynamic limit, the integral can be calculated by standard saddle-point
techniques. Thermodynamic equilibrium is then associated with the minimization
of $\psi$ with respect to $\left\{  q_{\alpha}\right\}  $, from which we
obtain self-consistent mean-field equations for these quantities. These
equations show that $q_{1}+q_{2}+q_{3}=0$, which suggests the introduction of
a symmetric traceless tensor,
\begin{equation}
\mathbf{Q}=\left(
\begin{array}
[c]{ccc}%
q_{1} & 0 & 0\\
0 & q_{2} & 0\\
0 & 0 & q_{3}%
\end{array}
\right)  ,
\end{equation}
as an appropriate thermal average of $\mathbf{\Omega}_{i}(n_{i})$. Using the
traceless condition, it is convenient to rewrite $\mathbf{Q}$ as
\begin{equation}
\mathbf{Q}=\frac{1}{2}\left(
\begin{array}
[c]{ccc}%
-S-\eta & 0 & 0\\
0 & -S+\eta & 0\\
0 & 0 & 2S
\end{array}
\right)  , \label{tensorQ}%
\end{equation}
in terms of two scalar parameters, $S$ and $\eta$. The isotropic phase is
given by $S=\eta=0$. The nematic uniaxial phase is given by $S\neq0$ and
$\eta=0$ (or $\eta=\pm3S$). In the biaxial phase, we have $S\neq0$ and
$\eta\neq0$.

In the following paragraphs we write explicit expressions for the
thermodynamic potentials of the uniform system and of the annealed binary
mixture. From these expressions, it is easy to perform numerical calculations
to draw a plethora of global phase diagrams in terms of the model parameters.
In order to asymptotically check the numerical findings, and to make contact
with established phenomenological results, we may also write an expansion of
the thermodynamic potential in terms of the invariants of the tensor order
parameter, $I_{n}=\text{Tr}\,\mathbf{Q}^{n}$, with $n=1$, $2$, $...$. Due to
the symmetry properties of $\mathbf{Q}$, all these invariants can be written
as polynomials depending on two basic invariants, given by
\begin{equation}
I_{2}=\text{Tr}\,\mathbf{Q}^{2}=\frac{1}{2}\left(  3S^{2}+\eta^{2}\right)  ,
\label{I2_trig}%
\end{equation}
and
\begin{equation}
I_{3}=\text{Tr}\,\mathbf{Q}^{3}=\frac{3}{4}S\left(  S^{2}-\eta^{2}\right)  .
\label{I3_trig}%
\end{equation}
Therefore, the usual form of the Landau--de Gennes expansion is written as
\begin{equation}
g=g_{0}+\frac{A}{2}\,I_{2}+\frac{B}{3}\,I_{3}+\frac{C}{4}\,I_{2}^{2}+\frac
{D}{5}\,I_{2}I_{3}+\frac{E}{6}\,I_{2}^{3}+\frac{E^{\prime}}{6}\,I_{3}^{2}+...
\label{Landaugen}%
\end{equation}
According to this phenomenological expansion \cite{Gramsbergen}, there is a
Landau multicritical point for $A=B=0$. In the vicinity of this Landau point,
we can establish parametric expressions for the lines of phase transitions
between the isotropic and nematic phases.

We now consider the specific cases of uniform and annealed systems.

\subsection{Uniform case}

Consider a system of intrinsically biaxial molecules ($\Delta_{i}=\Delta$ for
all $i$). Assuming the discretization of orientations, and setting $\mu=0$ in
Eq. (\ref{grandcanonical}), since disorder plays no role, the functional
$\psi$ is written as%
\begin{align}
\psi &  =\frac{1}{2}\left(  3S^{2}+\eta\right)  -T\ln2-T\ln\,{\LARGE \{}%
\,\exp\left[  -\frac{3\left(  S+\eta\right)  }{2T}\right]  \cosh\left[
\frac{\left(  -3S+\eta\right)  \Delta}{2T}\right]  +\label{funcionalbiaxial}\\
&  +\exp\left[  -\frac{3\left(  -S+\eta\right)  }{2T}\right]  \cosh\left[
\frac{\left(  3S+\eta\right)  \Delta}{2T}\right]  +\exp\left(  \frac{3S}%
{T}\right)  \cosh\left(  \frac{\eta\Delta}{T}\right)  \,{\LARGE \}},\nonumber
\end{align}
where $T=(\beta\epsilon)^{-1}$. Minimizing $\psi$ with respect to $S$ and
$\eta$ leads to the self-consistent mean-field equations, $S=F_{1}\left(
S,\eta;T,\Delta\right)  $ and $\eta=F_{2}\left(  S,\eta;T,\Delta\right)  $.
The values of $S$ and $\eta$ at the absolute minimum of $\psi$ correspond to
the thermodynamic equilibrium values for a fixed temperature $T $ and degree
of biaxiality $\Delta$. The free energy $f=f\left(  T,\Delta\right)  $ of the
system is obtained from $\psi$ by inserting the equilibrium values of $S$ and
$\eta$.

\begin{figure}[ptb]
\centering
\includegraphics[scale=0.45]{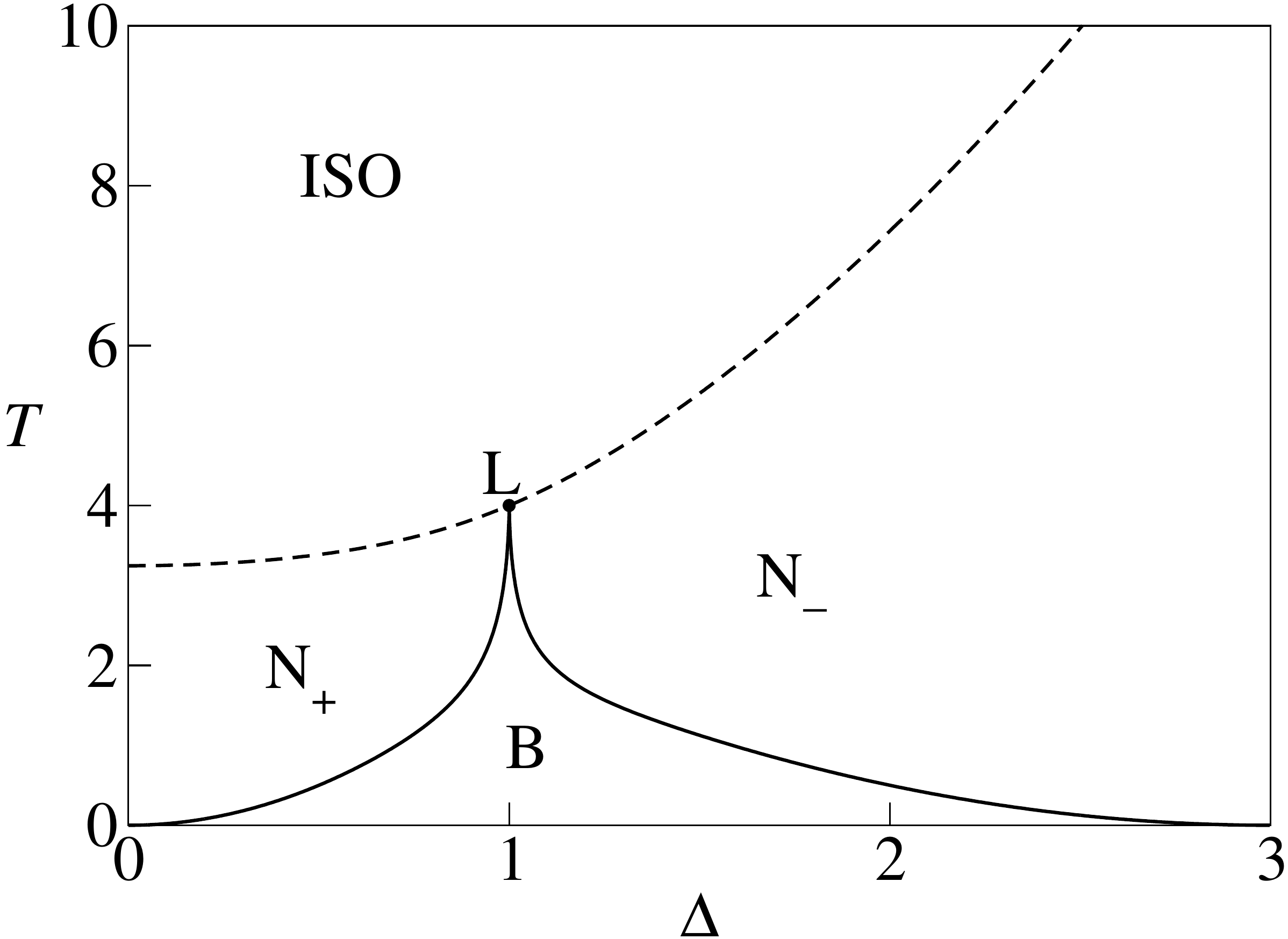}\caption{
Phase diagram, in terms of temperature $T$ and the degree of biaxiality
$\Delta$, for a system of intrinsically biaxial molecules. N$_{+} $ and
N$_{-}$ are uniaxial nematic prolate and oblate phases, respectively. B is the
nematic biaxial phase, L is the Landau multicritical point, and ISO is the
isotropic region. Continuous lines indicate continuous transitions and dashed
lines represent first-order phase transitions.}%
\label{phasediauniform}%
\end{figure}

Figure \ref{phasediauniform} shows the phase diagram in the $T-\Delta$ plane,
which is obtained by solving the mean-field equations numerically. As it
should be anticipated from phenomenological arguments, this phase diagram
shows two lines of continuous transitions (solid lines) from a biaxial nematic
region to the N$_{+}$ (prolate) and N$_{-}$ (oblate) uniaxial nematic regions.
These critical lines meet at a Landau multicritical point, L, on the
first-order boundary (dashed lines) between the isotropic and the uniaxial
nematic phases. It should be remarked that we regain an intrinsically uniaxial
system for $\Delta=3$. The phase diagram for $\Delta>3$ is mapped onto the
region $1<\Delta^{\prime}<3$ by the transformations $\Delta^{\prime}=\left(
\Delta+3\right)  /\left(  \Delta-1\right)  $ and $T^{\prime}=4T/\left(
\Delta-1\right)  ^{2}$. Note that a similar model for asymmetric ellipsoids
leads to essentially the same type of phase diagram \cite{Bocarra1970}. Also,
a number of calculations for continuous orientational degrees of freedom lead
to the same characteristic topology of this phase diagram (see, for example,
the works of Luckhurst and collaborators \cite{Luckhurst2012} and of Xheng and
Palffy-Murhoray \cite{Zheng2011}).

From the expression of the free energy, we obtain the parameter-dependent
coefficients of a Landau--de Gennes expansion about the Landau multicritical
point,
\begin{equation}
A=1-\frac{3+\Delta^{2}}{T};\qquad B=\frac{9}{2}\left(  \frac{\Delta^{2}%
-1}{T^{2}}\right)  ;
\end{equation}%
\begin{equation}
C=\frac{1}{T^{3}}\left(  \frac{9}{4}+\frac{3}{2}\Delta^{2}+\frac{1}{4}%
\Delta^{4}\right)  ,
\end{equation}%
\begin{equation}
D=-\frac{45}{16T^{4}}\left(  \Delta^{4}+2\Delta^{2}-3\right)  ,
\end{equation}%
\begin{equation}
E=-\frac{1}{480T^{5}}\left(  41\Delta^{6}+315\Delta^{4}+1215\Delta
^{2}+1053\right)  ,
\end{equation}%
\begin{equation}
E^{\prime}=\frac{1}{40T^{5}}\left(  \Delta^{6}+225\Delta^{4}-405\Delta
^{2}+243\right)  .
\end{equation}
Therefore, the Landau point L is located at $T_{L}=4$ and $\Delta_{L}=1$. In
the vicinity of L, limiting to first order terms in $\left(  T-4\right)  $ and
$\left(  \Delta-1\right)  $, we have $A=\left(  1/4\right)  \left(
T-4\right)  -\left(  1/2\right)  (\Delta-1)$, $B=\left(  9/16\right)  \left(
\Delta-1\right)  $, $C=1/16$, $D=0$, $E=-41/7680$, and $E^{\prime}=1/640$. The
sign of $E^{\prime}$ indicates the stability of the biaxial nematic phase near
the Landau point \cite{deGennes, Gramsbergen}. In this mean-field
scenario, at fixed $\Delta\neq1$, as the temperature decreases from a
sufficiently large value, the system goes from an isotropic phase to a
uniaxial nematic phase, and then to a biaxial nematic phase, according to the
prediction of the early work of Freiser \cite{Freiser}. It should be remarked
that the phenomenological Landau parameters are written in terms of the
parameters of the underlying molecular model, which makes it easier to
investigate a large range of values.

\subsection{Annealed disorder}

In the annealed case, we calculate $\psi$, given by Eq. (\ref{grandpot}), in
terms of the proper thermodynamic field variables, temperature $T$ and
chemical potential $\mu=\epsilon T\ln z$, where $z$ is a fugacity. For the
fully-connected MS6 model of a binary mixture of uniaxial and biaxial
molecules, the functional $\psi$ is given by%
\begin{equation}
\psi=\frac{1}{2}\left(  3S^{2}+\eta^{2}\right)  -T\ln2-T\ln\sigma,
\end{equation}
where%

\begin{equation}%
\begin{split}
\sigma &  =e^{-\frac{3\left(  S+\eta\right)  }{2T}}\left\{  1+z\cosh\left[
\frac{\left(  3S-\eta\right)  \Delta}{2T}\right]  \right\} \\
&  \quad+e^{\frac{-3\left(  S-\eta\right)  }{2T}}\left\{  1+z\cosh\left[
\frac{\left(  3S+\eta\right)  \Delta}{2T}\right]  \right\} \\
&  \quad+e^{\frac{3S}{T}}\left\{  1+z\cosh\left(  \frac{\eta\Delta}{T}\right)
\right\}  \quad.
\end{split}
\end{equation}
Again, the thermodynamic stable values of $S$ and $\eta$ are chosen to
minimize the function $\psi$ for fixed values of temperature $T$, degree of
biaxiality $\Delta$, and chemical potential $\mu$.

Inserting the equilibrium values of $S$ and $\eta$ into the expression of
$\psi$, we obtain the grand potential as a function of $T$, $\mu$, and
$\Delta$. If we wish to work with a fixed concentration $c$ of the
intrinsically biaxial molecules, the free energy $\phi\left(  c,T;\Delta
\right)  $ comes from the definition%
\begin{equation}
\phi\left(  c,T;\Delta\right)  =\psi+c\ln z,
\end{equation}
where the fugacity is eliminated by the expression%
\begin{equation}
c=-z\frac{\partial\psi}{\partial z},
\end{equation}
and we should insert the equilibrium values of $S$ and $\eta$ (with the
proviso of a Maxwell construction whenever it is necessary).

The grand potential can be used to write a Landau--de Gennes expansion with
coefficients
\begin{equation}
A=1-\frac{3+z\left(  3+\Delta^{2}\right)  }{T\left(  1+z\right)  };\qquad
B=\frac{9}{2}\frac{z\left(  \Delta^{2}-1\right)  -1}{T^{2}\left(  1+z\right)
};
\end{equation}%
\begin{equation}
C=\frac{9+z\left[  18+9z+6(1+z)\Delta^{2}-(1-z)\Delta^{4}\right]  }%
{4T^{3}\left(  1+z\right)  ^{2}};
\end{equation}%
\begin{equation}
D=\frac{15}{16}\frac{9+z\left[  18-6\Delta^{2}+\Delta^{4}-3z\left(
-3+2\Delta^{2}+\Delta^{4}\right)  \right]  }{T^{4}\left(  1+z\right)  ^{2}};
\end{equation}%
\begin{equation}
E=\frac{1053-z\left(  c_{0}+zc_{1}+z^{2}c_{2}\right)  }{480T^{5}\left(
1+z\right)  ^{3}},
\end{equation}
where
\begin{equation}
c_{0}=-3159-1215\Delta^{2}+225\Delta^{4}-11\Delta^{6},
\end{equation}%
\begin{equation}
c_{1}=-3159-2430\Delta^{2}-90\Delta^{4}+68\Delta^{6}%
\end{equation}
and
\begin{equation}
c_{2}=-1053-1215\Delta^{2}-315\Delta^{4}-41\Delta^{6},
\end{equation}%
\begin{equation}
E^{\prime}=\frac{243+z\left(  d_{0}+zd_{1}\right)  }{40T^{5}\left(
1+z\right)  ^{2}},
\end{equation}
where
\begin{equation}
d_{0}=486-405\Delta^{2}-45\Delta^{4}+\Delta^{6},
\end{equation}
and
\begin{equation}
d_{1}=243-405\Delta^{2}+225\Delta^{4}+\Delta^{6}.
\end{equation}
From the usual condition $A=B=0$, we locate a Landau multicritical point,%
\begin{equation}
\mu_{L}=-T_{L}\ln\left(  \Delta^{2}-1\right)  ,\quad T_{L}=4,
\end{equation}
which requires $\Delta^{2}>1$, and gives an indication of the existence of
qualitatively different phase diagrams (as there is no Landau point for
$\Delta^{2}<1$).

\begin{figure}[ptb]
\caption{ Energy levels as a function of the degree of biaxiality $\Delta$.
Level crossings account for variations in the entropic contribution to the
free energy. }%
\label{energylevels}
\centering
\includegraphics[scale=0.45]{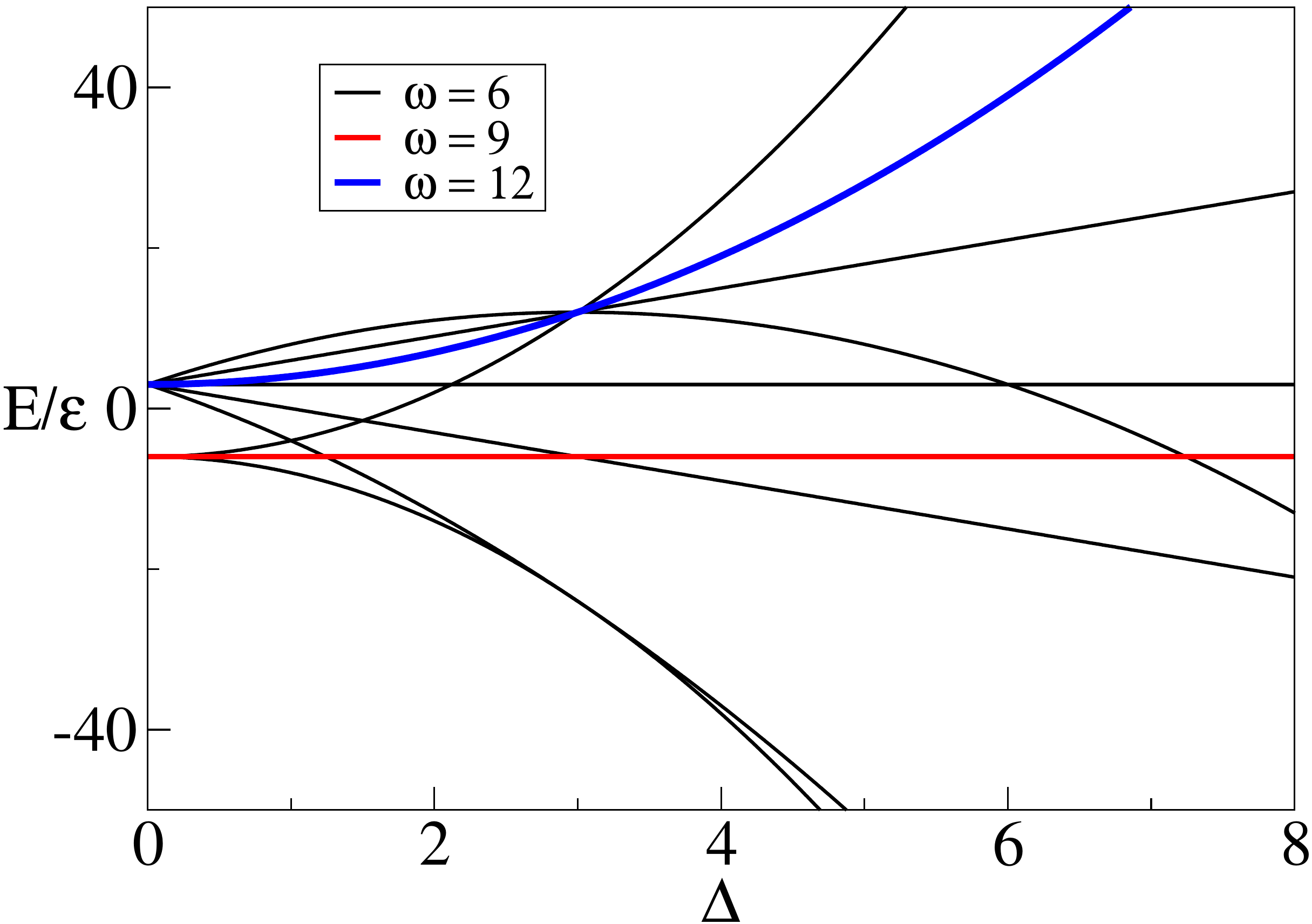}\end{figure}

It is easy to show that this annealed version of the binary mixture exhibits
phase diagrams with many distinct topologies. This can be anticipated from an
analysis of the energy levels associated with the interaction between
molecules, as shown in Figure \ref{energylevels}. In fact, these energy levels
are associated with different degrees of degeneracy $\omega$, and some levels
cross each other as the biaxiality of the molecules is changed. These
degeneracies account for entropic contributions to the free energy, which do
affect the equilibrium phase behavior of the system. Therefore, we anticipate
qualitative changes in the phase diagrams as $\Delta$ assumes values close to
the location of the energy level crossings.

We now discuss the various topologies exhibited by the phase diagrams as the
biaxiality parameter is changed. 

\begin{figure}[ptb]
\centering
\includegraphics[scale=0.45]{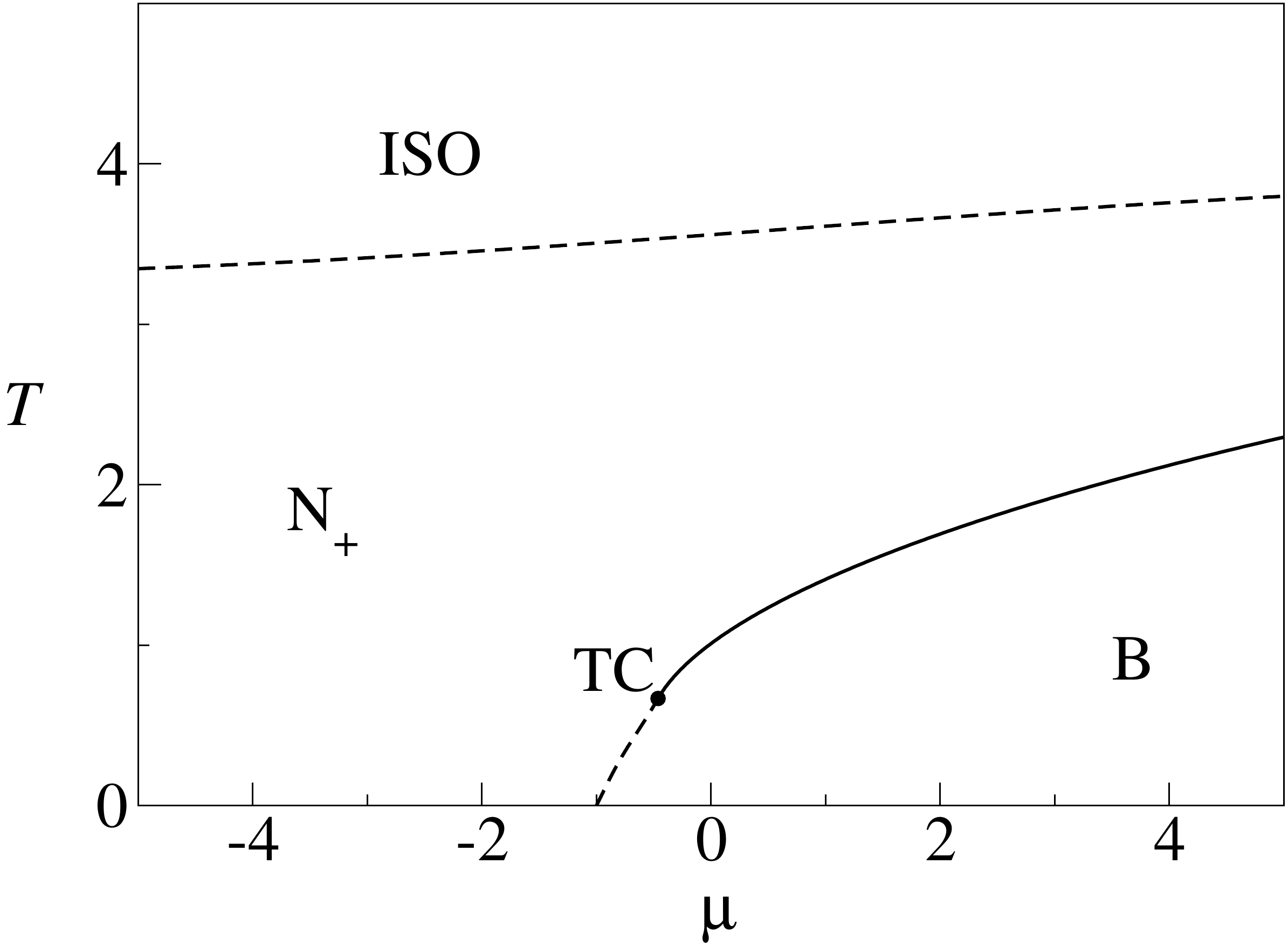}\caption{
Phase diagram for fixed degree of biaxiality $\Delta=1$, where $T$ is
temperature and $\mu$ is chemical potential. There is a tricritical point, TC,
along the boundary separating biaxial (B) and uniaxial prolate nematic
(N$_{+}$) phases. There is no direct phase transition between the isotropic
(ISO) and biaxial phases (for finite values of the chemical potential). }%
\label{muvsT4Delta=1}%
\end{figure}

\begin{figure}[ptb]
\centering
\includegraphics[scale=0.45]{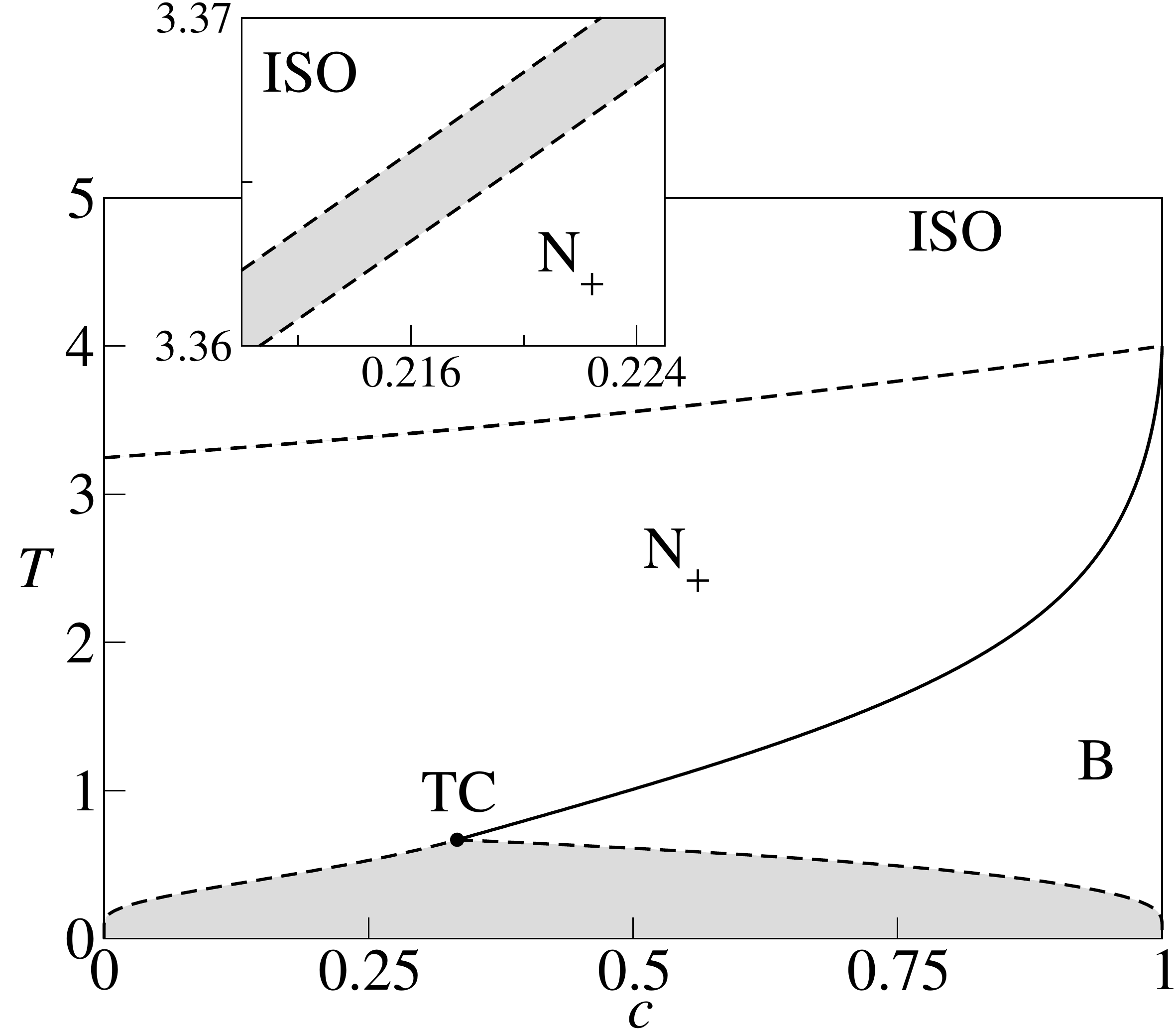}\caption{
Phase diagram in terms of the temperature ($T$) versus the concentraton ($c$)
of biaxial molecules, at a fixed degree of biaxiality, $\Delta=1$. There is an
incipient Landau multicritical point at $c=1$, where the two ordered phases
and the isotropic phase become identical. Uniaxial (N$_{+}$) and biaxial (B)
phases coexist at low temperatures and intermediate concentrations (gray
region). The system exhibits a tricritical point, TC, along the boundary of
the B phase. The inset represents a zoom of the coexistence (gray) region
between the uniaxial nematic and isotropic phases. }%
\label{convsT4Delta=1.0}%
\end{figure}

Figures \ref{muvsT4Delta=1}-\ref{convsT4Delta=1.0} show $T-\mu$ and $T-c$
phase diagrams for fixed degree of biaxiality, $\Delta=1$, where $\mu$ is the
chemical potential and $c$ is the concentration of intrinsically biaxial
molecules. From the thermodynamic point of view, the first-order boundaries
(dashed lines) in the $T-\mu$ plane are mapped into coexistence regions (gray
regions) in the $T-c$ plane. For low temperatures and intermediate
concentrations, there is a coexistence region between the uniaxial prolate
(N$_{+}$) and biaxial (B) phases. However, at higher concentrations and
intermediate temperatures, there is a second-order phase transition
(continuous line) between the N$_{+}$ and B phases. In fact, the phase diagram
exhibits a tricritical point, TC, along the boundary between N$_{+}$ and B. At
higher temperatures, there is a first-order phase transition between N$_{+}$
and the isotropic (ISO) phases, with a very thin coexistence region, as shown
in the inset. Note that an incipient Landau point appears at $c=1$, which
corresponds to an infinite chemical potential, in agreement with the
Landau--de Gennes expansion. Phase diagrams with a similar topology (but with
no Landau point) can be drawn for $0<\Delta<1$.

\begin{figure}[ptb]
\centering
\includegraphics[scale=0.45]{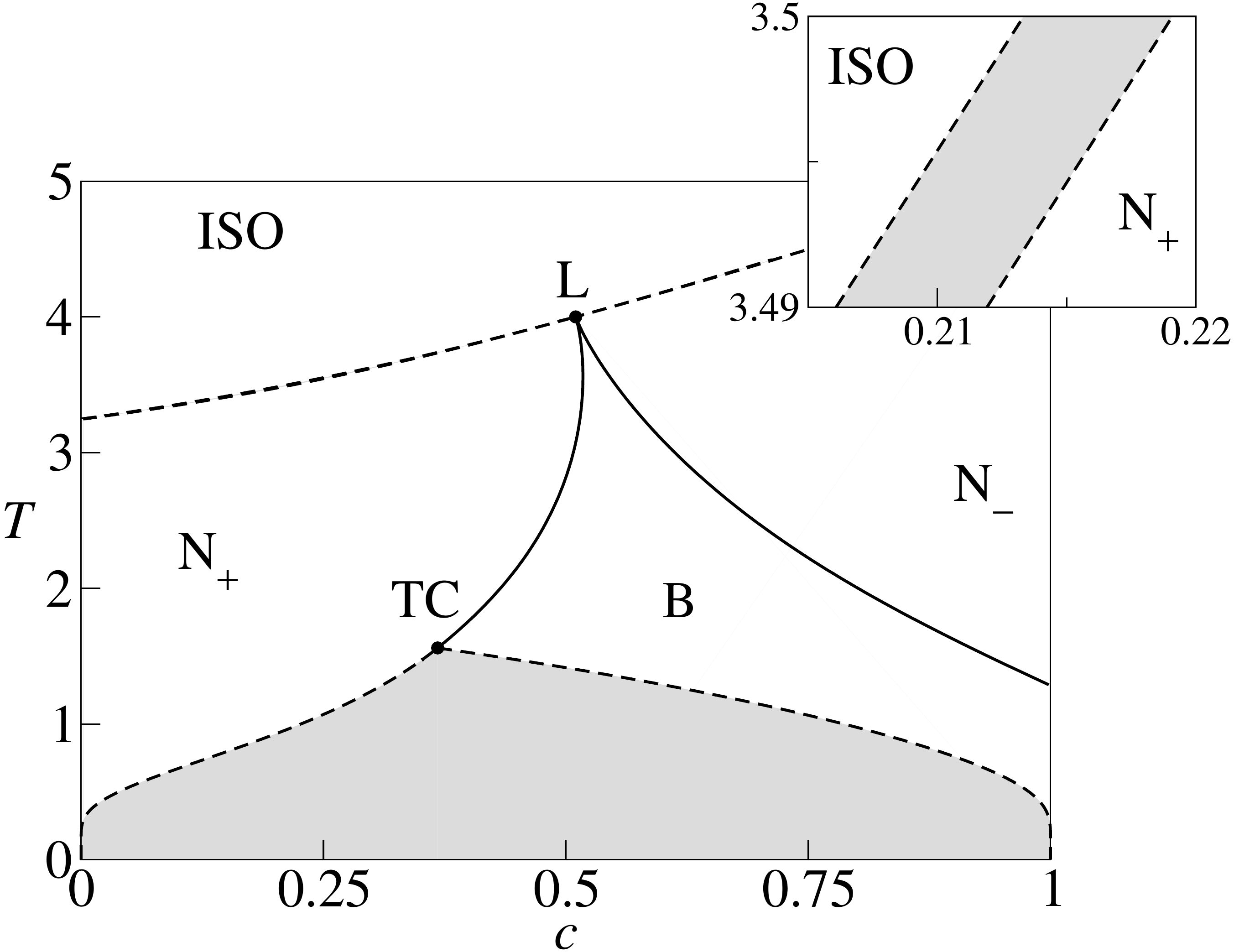}\caption{
Phase diagram, in terms of temperature $T$ and concentration $c$ of biaxial
molecules, for degree of biaxiality $\Delta=1.4$. There is a Landau point L.
Note that the biaxial phase displays a reentrant behavior near L. }%
\label{convsT4Delta=1.4}%
\end{figure}

Figure \ref{convsT4Delta=1.4} shows the phase diagram for a fixed degree of
biaxiality $\Delta=1.4$. Similar to Figure \ref{convsT4Delta=1.0}, the N$_{+}$
and B phases coexist for intermediate concentrations and low temperatures.
However, the system also displays a uniaxial oblate (N$_{-}$) nematic phase,
which appears at higher concentrations and intermediate temperatures. The B
phase appears between the two uniaxial phases, at intermediate temperatures.
All three ordered phases become identical to the ISO phase at a Landau
multicritical point, L. Also, note that the B phase presents a discrete
reentrant behavior close to L. Note that the changes in the topology of the
phase diagrams shown in Figures \ref{convsT4Delta=1.0} and
\ref{convsT4Delta=1.4} are in agreement with the dependence of the energy
levels on the degree of biaxiality $\Delta$ (see Figure \ref{energylevels}).

\begin{figure}[ptb]
\centering
\includegraphics[scale=0.45]{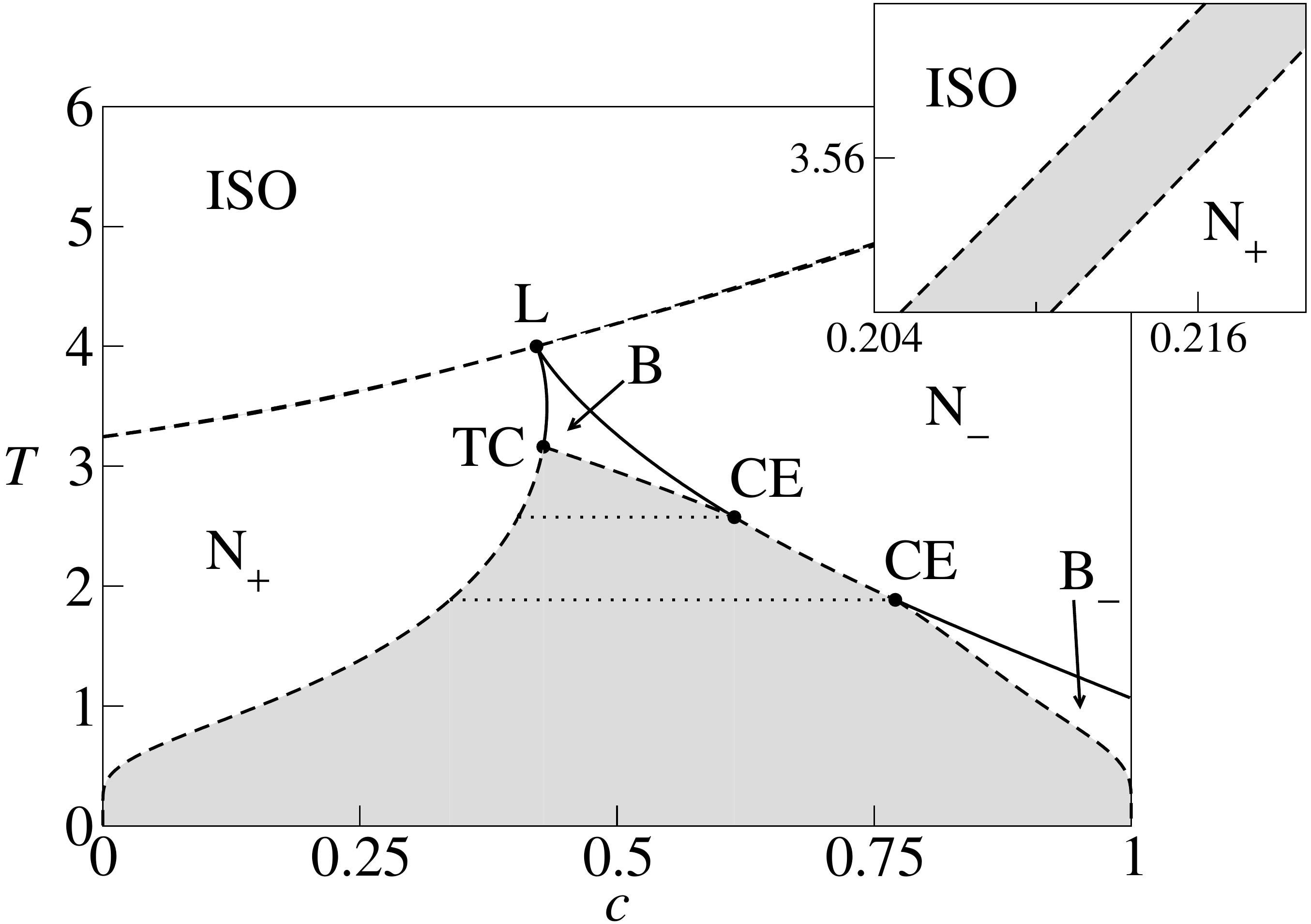}\caption{
Phase diagram in the $c-T$ plane for degree of biaxiality $\Delta=1.54$, where
$T$ is the temperature and $c$ is the concentation of biaxial objects. Two
nematic biaxial phases are stable in disconnected regions of the phase
diagram. There is a coexistence region between the nematic uniaxial phases,
ending at two critical end points CE. The dotted horizontal lines represent
special tie lines associated with the critical end points.}%
\label{convsT4Delta=1.54}%
\end{figure}

The phase behavior of the system changes significantly for degree of
biaxiality around $\Delta=1.5$, which is close to another crossing of energy
levels. For example, in Figure \ref{convsT4Delta=1.54} we show the phase
diagram for degree of biaxiality $\Delta=1.54$. The low temperature biaxial
phase B$_{-}$ is represented by a tensor order parameter $\mathbf{Q}$ whose
largest eigenvalue (in absolute value) is negative. However, an additional
biaxial phase B appears near the Landau point. The two biaxial phases are
stable in disconnected regions of the phase diagram. There is then a
coexistence region between the uniaxial nematic phases N$_{+}$ and N$_{-}$.
This region is limited by two critical end points, CE, associated with the
biaxial phases. Also, the system exhibits a tricritical point TC related to
the coexistence region of N$_{+}$ and B structures. As in the case
$\Delta=1.4$, shown in Figure \ref{convsT4Delta=1.4}, the biaxial phase is
reentrant in the vicinity of the Landau point.

\begin{figure}[ptb]
\centering
\includegraphics[scale=0.45]{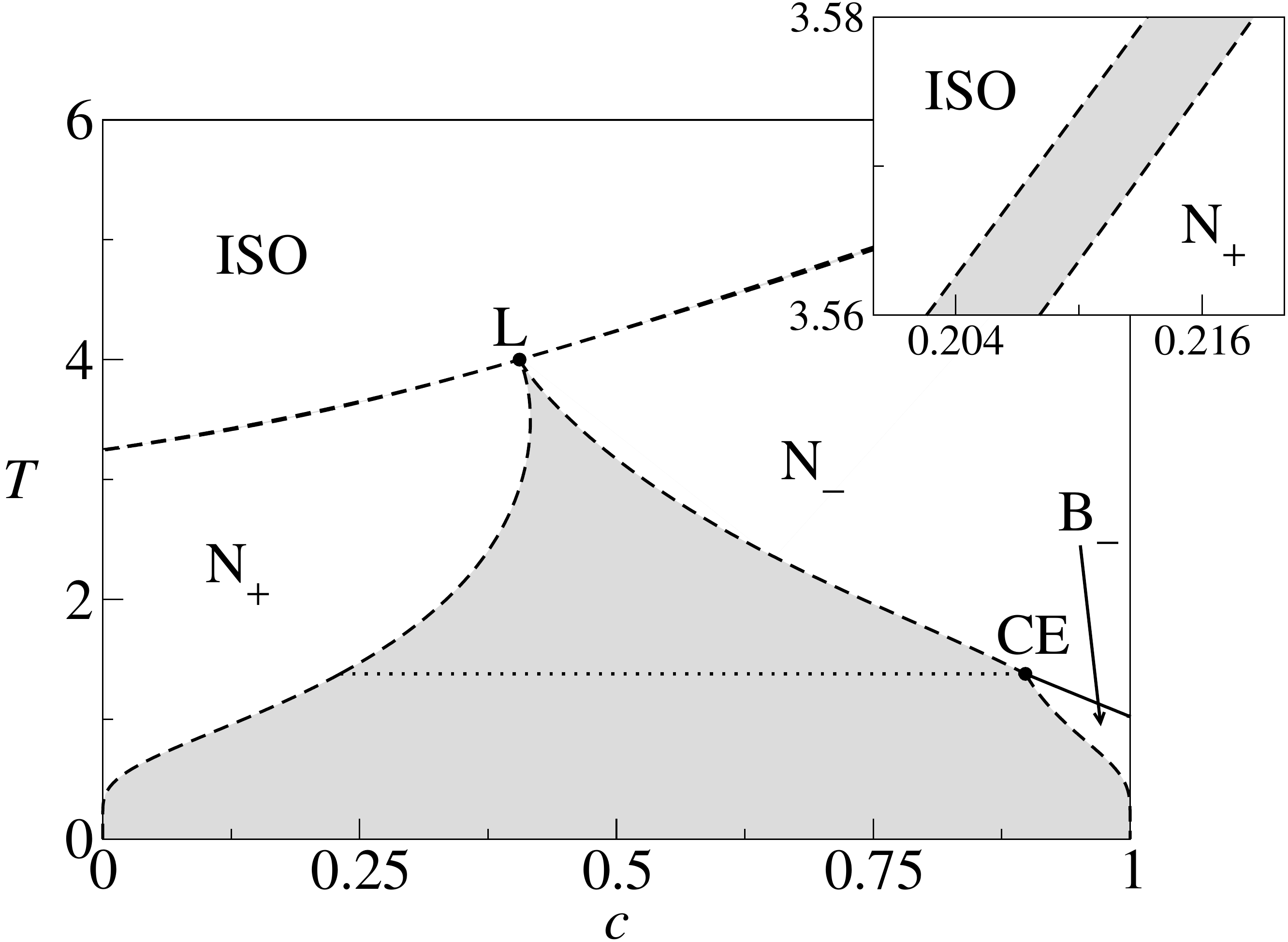}\caption{
Phase diagram in terms of temperature ($T$) versus concentrarion ($c$) of
biaxial molecules for fixed degree of biaxiality, $\Delta=1.57$. The biaxial
phase B$_{-}$ is located at low temperatures and high concentations. There is
no stable B phase near the Landau point L. The nematic uniaxial phases coexist
in a region ending at a critical end point CE. The dotted horizontal line
represents a special tie line associated with the critical end point.}%
\label{convsT4Delta=1.57}%
\end{figure}

\begin{figure}[ptb]
\centering
\includegraphics[scale=0.45]{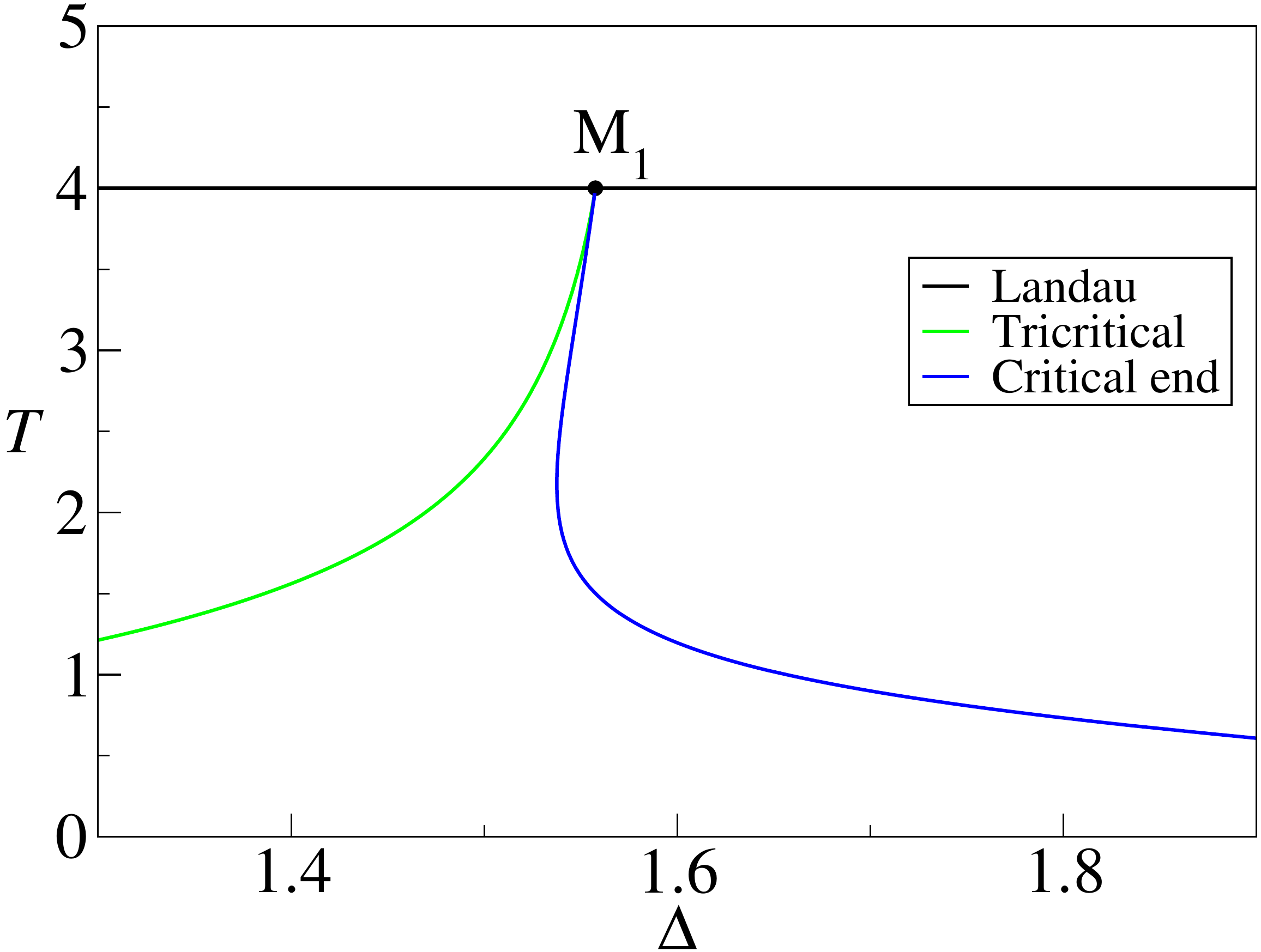}\caption{
Lines of multicritical points in the $\Delta-T$ plane. There is a higher-order
multicritical point M$_{1}$, at which the lines of Landau points (black),
tricritical points (green) and critical end points (blue) meet. There is then
no stable biaxial phase for $\Delta\geq\Delta_{\text{M}_{1}}$ in the vicinity
of the Landau point.}%
\label{DelvsTemp1}%
\end{figure}

The topological changes in the phase diagram can be represented by the
projections of the lines of different multicritical points on the $T-\Delta$
plane, as indicated in Figure \ref{DelvsTemp1}. The temperature of the Landau
point (black) is a constant function of the degree of biaxiality $\Delta$,
which is in agreement with a Landau expansion. The temperature of the
tricritical point (green) increases monotonically with $\Delta$, as suggested
by Figures \ref{convsT4Delta=1.0}-\ref{convsT4Delta=1.54}. However, the line
of critical end points (blue) presents a reentrant behavior in the vicinity of
$\Delta=1.55$, giving rise to the two critical end points, as it is shown in
Figure \ref{convsT4Delta=1.54}. All multicritical lines meet at a higher-order
multicritical point M$_{1}$. Note that the tricritical and high-temperature
critical end points are associated with the biaxial phase near the Landau
point. As a result, there is no stable biaxial phase in the vicinity of L for
$\Delta\geq\Delta_{M_{1}}\simeq1.5576$, although the low-temperature biaxial
phase survives in a small region at high concentrations and low temperatures.
This is illustrated in Figure \ref{convsT4Delta=1.57}, for $\Delta=1.57$.
There is no stable biaxial nematic phase close to the Landau point L, which
marks the meeting of various coexistence lines separating the isotropic phase
and the two nematic uniaxial phases. Along those lines, there is a coexistence
of phases with different values of nematic order parameter $S$, but the size
of the coexistence region tends to vanish as we approach the Landau point.

\begin{figure}[ptb]
\centering
\includegraphics[scale=0.45]{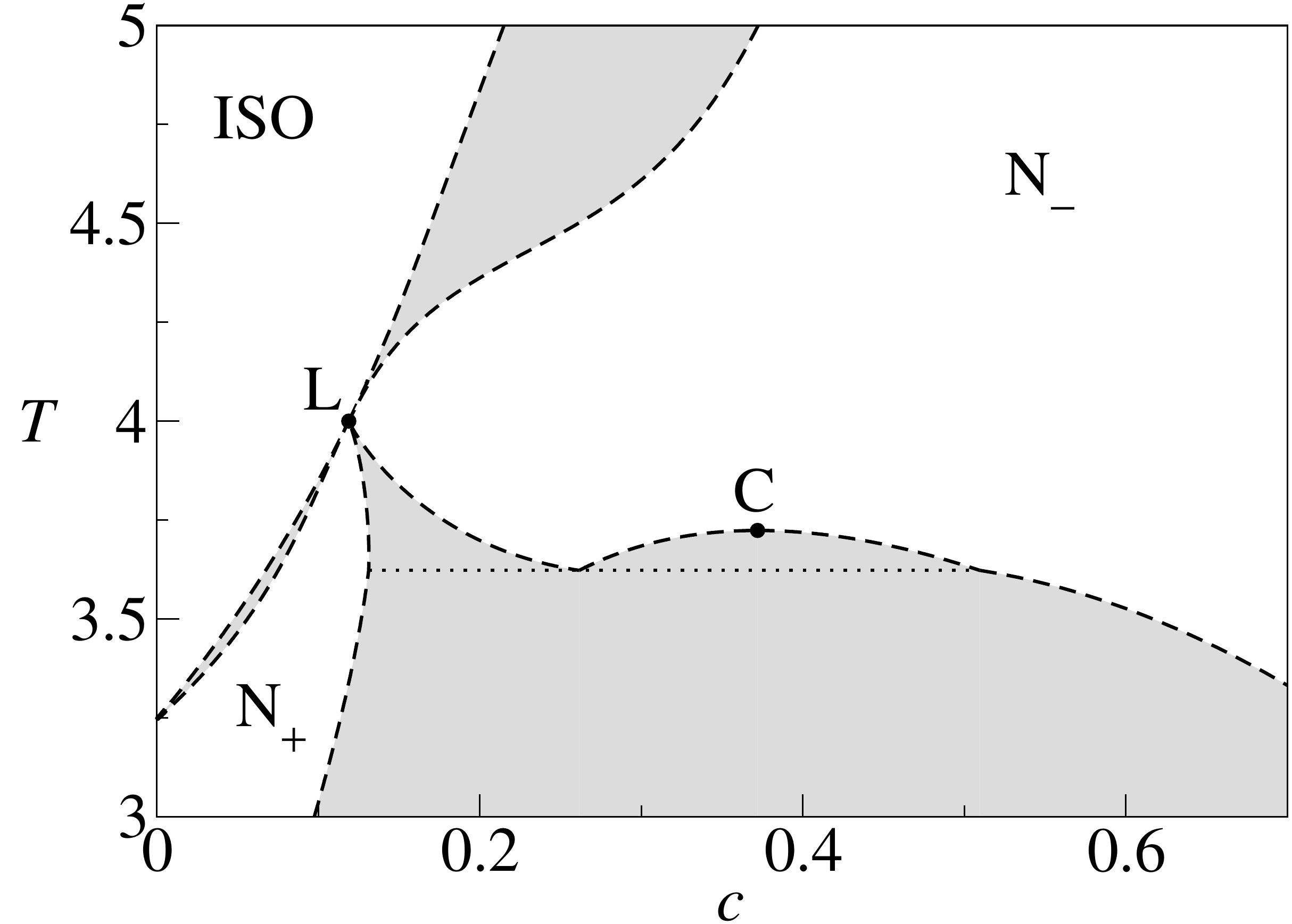}\caption{
Phase diagram in terms of temperature $T$ and concentration $c$ of biaxial
objects, for degree of biaxiality $\Delta=2.9$. There is no biaxial nematic
phase around the Landau point L. C indicates a simple critical point
associated with a coexistence region between two uniaxial oblate phases. The
dotted line indicates a triple point, corresponding to the coexistence of two
uniaxial oblate and one uniaxial prolate phases.}%
\label{convsT4Delta=2.9}%
\end{figure}

\begin{figure}[ptb]
\centering
\includegraphics[scale=0.45]{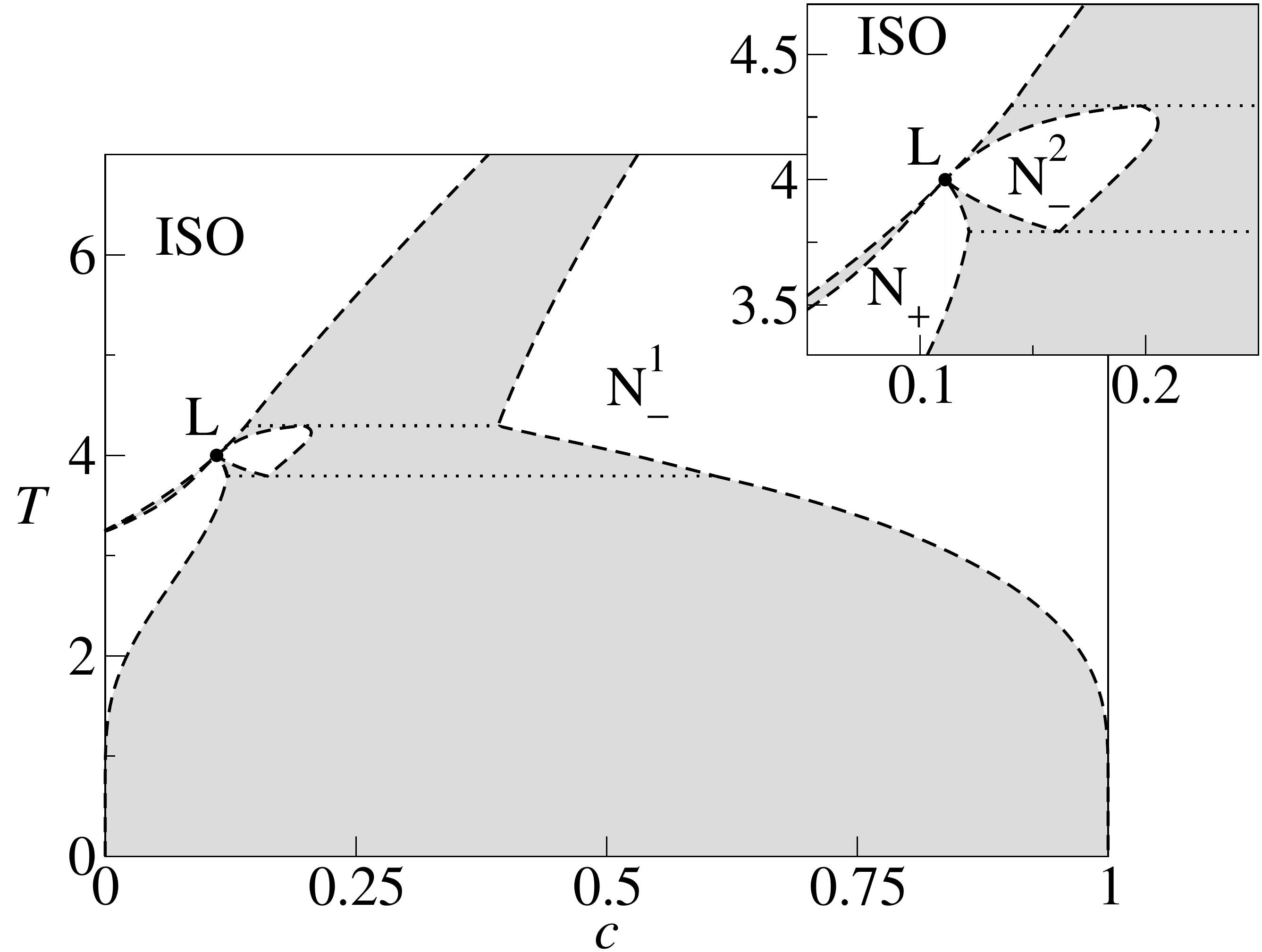}\caption{
Phase diagram in terms of temperature $T$ and concentration $c$ of biaxial
objects, for degree of biaxiality $\Delta=3.0$. This system corresponds to a
mixture of rods and plates with asymmetric interaction energies. A uniaxial
nematic prolate phase, $\mathrm{N}_{+}$, and two uniaxial nematic oblate
phases, $\mathrm{N}_{-}^{1}$ and $\mathrm{N}_{-}^{2}$, are present. There are
two triple points indicated by the dotted tie lines. }%
\label{convsT4Delta=3.0}%
\end{figure}

As the degree of biaxiality is further increased, the system exhibits other
distinct phase diagrams, as seen in Figures \ref{convsT4Delta=2.9}%
-\ref{convsT4Delta=3.0}. According to Figure \ref{energylevels}, multiple
level crossings occur at $\Delta=3$, which is a strong suggestion of changes
in the phase behavior of the system.

Figure \ref{convsT4Delta=2.9} shows the phase diagram for $\Delta=2.9$. There
appears a triple point associated with the coexistence of two uniaxial nematic
oblate phases and a uniaxial prolate phase. Also, the coexistence region of
uniaxial nematic oblate phases ends at a simple critical point C. Although it
is not shown in Figure \ref{convsT4Delta=2.9}, a stable biaxial nematic phase
is still present, at low temperatures and high concentrations, as well as the
critical end point associated with the biaxial phase.

As $\Delta$ is further increased, the simple critical point moves upward in
the phase diagram, approaching the lower border of the coexistence region
between the uniaxial $\mathrm{N}_{-}$ and ISO phases. Then, the simple
critical point is replaced by a second triple point. For example, Figure
\ref{convsT4Delta=3.0} represents a phase diagram for $\Delta=3$. Due to the
second triple point, there appears an isolated second uniaxial oblate phase
$\mathrm{N}_{-}^{(2)}$. Note that, according to Eq.(\ref{seisa}), the value
$\Delta=3$ corresponds to a binary mixture of rods and plates with asymmetric
interaction energies. A symmetric choice of interaction energies
\cite{ECarmo2011} presents a much simpler phase diagram, with single nematic
uniaxial prolate and oblate phases, in addition to an isotropic phase, and no
triple points.

\begin{figure}[ptb]
\centering
\includegraphics[scale=0.45]{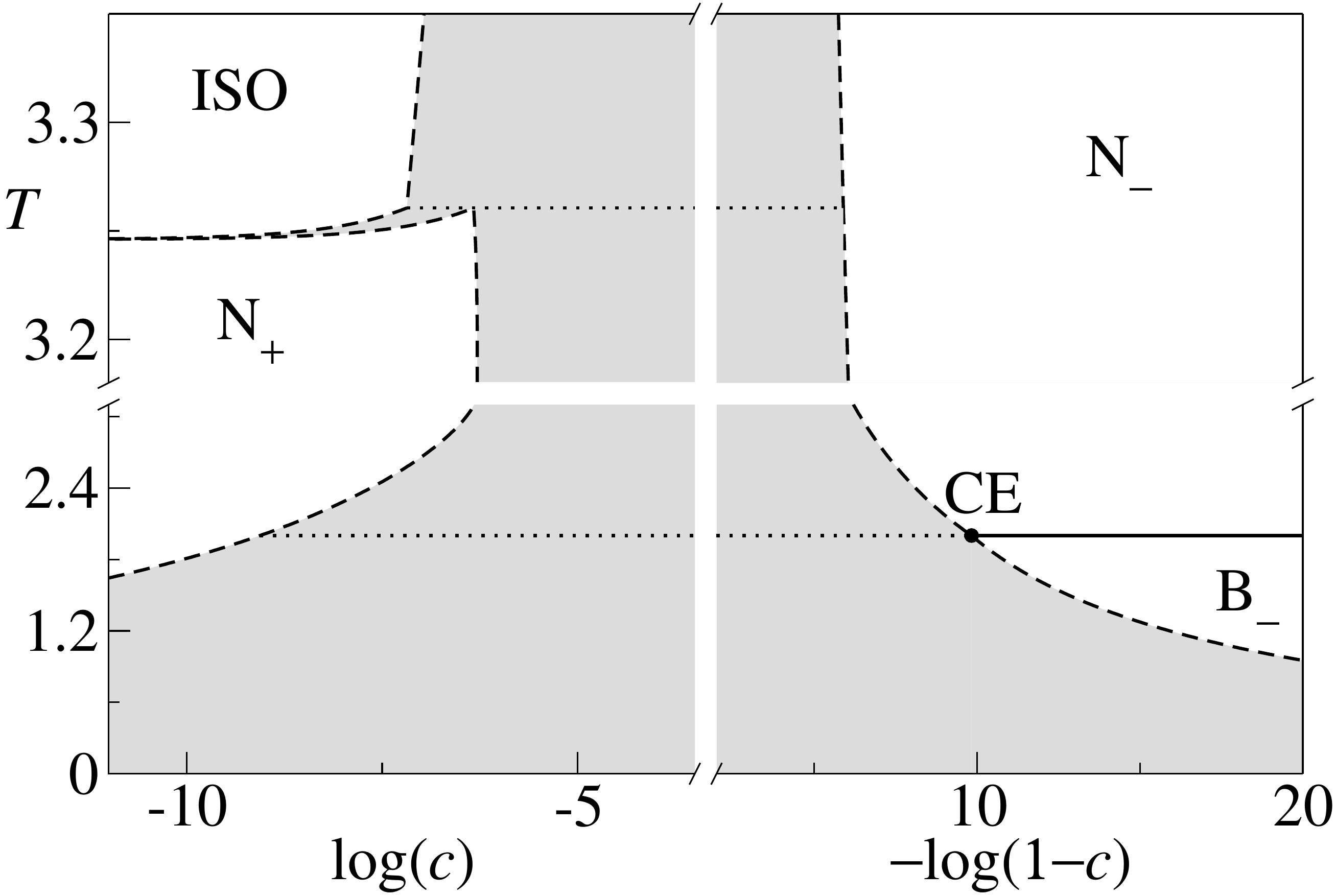}\caption{
Phase diagram in terms of temperature $T$ and concentration $c$ of biaxial
molecules, for degree of biaxiality $\Delta=5.0$. The Landau point is absent.
There appears a triple point, at which N$_{+}$, N$_{-}$ and ISO phases
coexist. The B$_{-}$ phase is stable at high concentations. }%
\label{convsT4Delta=5.0}%
\end{figure}

\begin{figure}[ptb]
\centering
\includegraphics[scale=0.45]{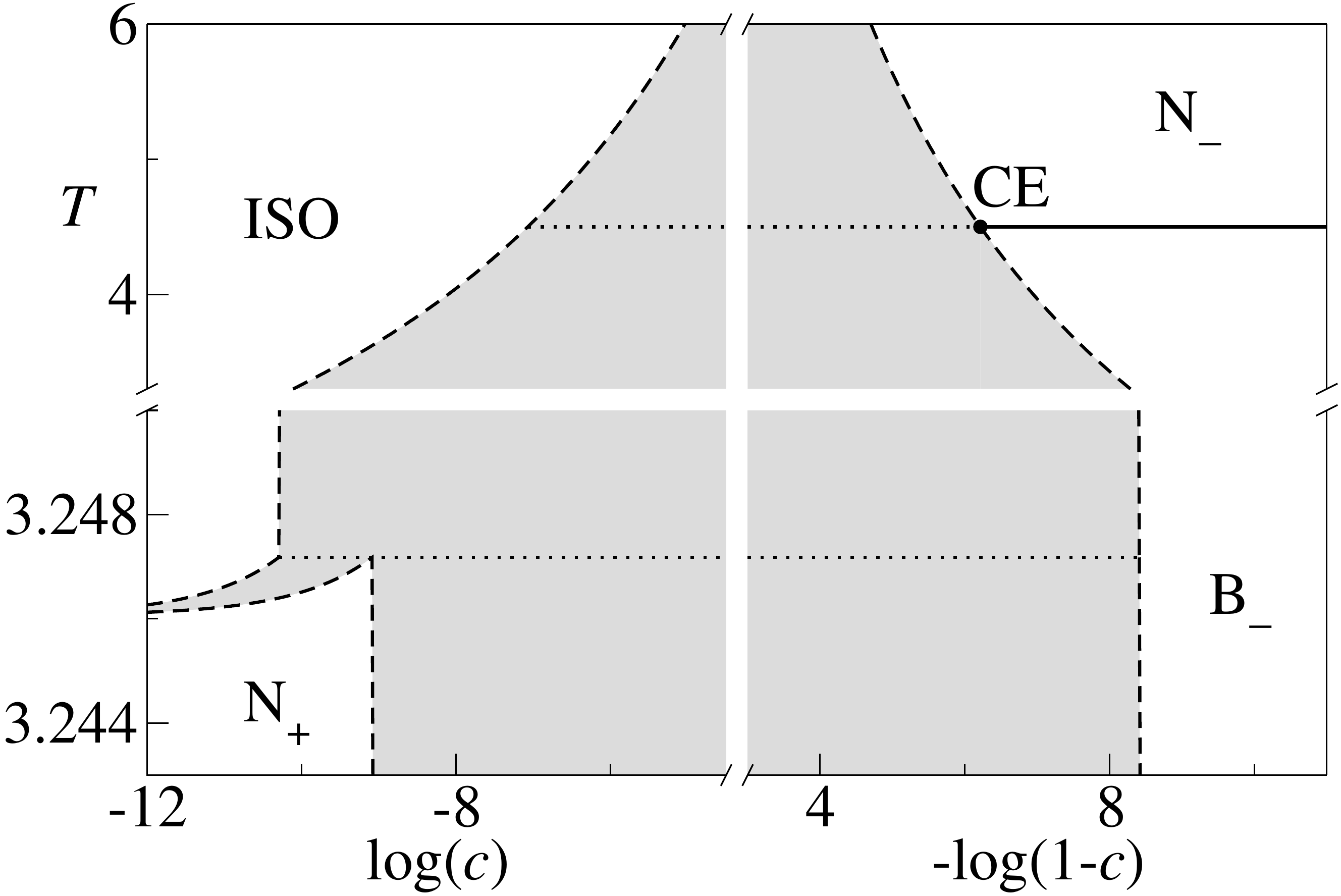}\caption{
Phase diagram in terms of temperature $T$ and concentration $c$ of biaxial
objects, for degree of biaxiality $\Delta=6.0$. The Landau point is absent.
There is a coexistence region between ISO and B$_{-}$ phases. In addition,
ISO, B$_{-}$ and N$_{+}$ phases coexist at a triple point. The B$_{-}$ phase
is stable at high concentations. }%
\label{convsT4Delta=6.0}%
\end{figure}

Figures \ref{convsT4Delta=5.0} and \ref{convsT4Delta=6.0} represent the phase
diagrams for $\Delta=5$ and $\Delta=6$, respectively. In these diagrams the
values of concentrations are conveniently rescaled, so that visual effects are
improved. Also, note that we introduce some separations just to emphasize the
more interesting sectors of this phase diagram. There is no stable Landau
point in both phase diagrams. In Figure \ref{convsT4Delta=5.0}, the uniaxial
phases N$_{+}$ and N$_{-}$ coexist with the ISO phase at a triple point, and a
biaxial B$_{-}$ phase remains stable at high concentrations. However, in
Figure \ref{convsT4Delta=6.0}, there is a coexistence region of B$_{-}$ and
ISO phases, as well as a triple point of coexistence of ISO, B$_{-}$ and
N$_{+}$ phases. Figures \ref{convsT4Delta=5.0} and \ref{convsT4Delta=6.0}
suggest that the triple points meet the critical end point as the degree of
biaxiality is increased.

\begin{figure}[ptb]
\centering
\includegraphics[scale=0.45]{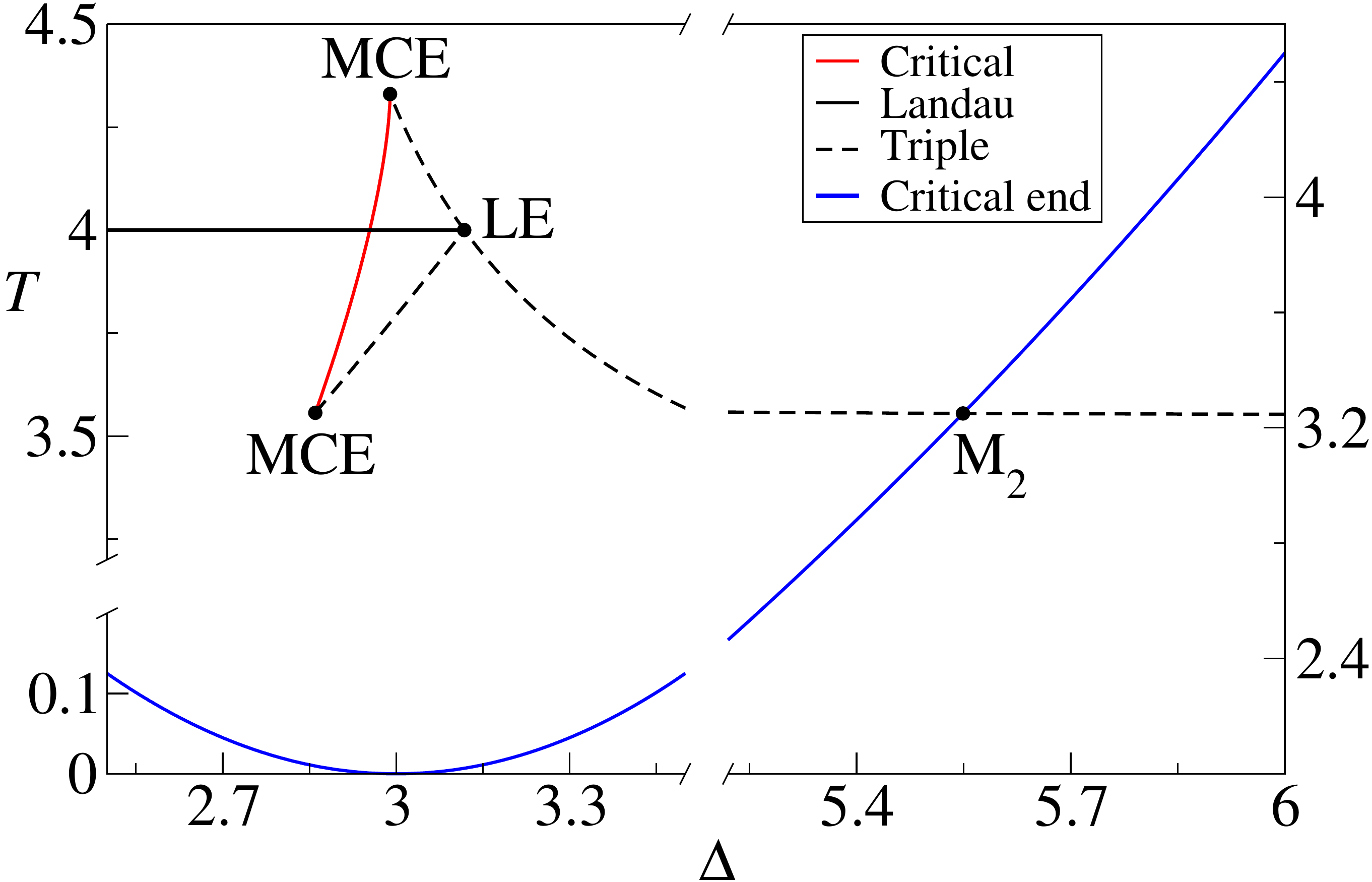}\caption{Lines
of phase transitions in the $\Delta-T$ plane. The dashed black lines represent
lines of triple points. The line of simple critical points (red) ends at two
multicritical end points MCE. The line of Landau poins meets the lines of
triple points at a Landau end point LE. Also, the line of critical end points
(blue) crosses the line of triples points at a multicritical point M$_{2}$. }%
\label{DelvsTemp2}%
\end{figure}

In Figure \ref{DelvsTemp2} we draw the projections of distinct multicritical
points on the $\Delta-T$ plane. The line of Landau points is still present, as
well as the low-temperature part of the line of critical end points associated
with B$_{-}$. Furthermore, there are lines of triple points associated with
the ISO and various nematic uniaxial phases, as depicted in Figures
\ref{convsT4Delta=2.9} and \ref{convsT4Delta=3.0}. The line of Landau points
meets the triple lines at another special multicritical point, which we call a
Landau end point, LE. This point is characterized by the coexistence of a
critical ISO phase and a non-critical N$_{-}$ phase. Otherwise, two triple
lines meet the line of simple critical points at two multicritical end points,
MCE. In addition, for $\Delta\simeq5.5$, the triple line meets the critical
end line at a multicritical point, M$_{2}$, where the critical phase N$_{-}$
coexists with the non-critical phases N$_{+} $ and ISO. Consequently, for
$\Delta\geq5.5$, there is a coexistence region between ISO and B$_{-}$ phases
as we increase the concentration of biaxial molecules.

\section{Conclusions}

\label{sec:conc}

We introduced an elementary six-state Maier-Saupe (MS6) lattice model, which
is obtained by the addition of extra degrees of freedom, of biaxial nature, to
an earlier three-state model. We then described a phase diagram with biaxial
as well uniaxial nematic structures and an isotropic phase. A fully connected
MS6 lattice model, of mean-field character, is sufficiently simple to be
amenable to a detailed treatment by standard statistical mechanics techniques.
Results are obtained in terms of a parameter $\Delta$ of shape fluctuations
that gauges the degree of biaxiality. We then used this MS6 model to consider
a binary mixture of intrinsically uniaxial ($\Delta=0$) and intrinsically
biaxial ($\Delta\neq0$) molecules, and investigated the effects of
\textquotedblleft shape fluctuations" on the phase diagrams in terms of
temperature and either chemical potential or concentration of biaxial molecules.

Taking into account the fluid character of the liquid-crystalline systems, in
the present work we restricted the analysis to a thermalized (annealed)
situation, in which case orientational and shape degrees of freedom reach
equilibrium simultaneously. We obtained a wealth of topologically distinct
phase diagrams, with several nematically ordered structures and multicritical points.

In the uniform case, in terms of temperature $T$ and the parameter $\Delta$,
we regained the first-order transitions between isotropic and uniaxial nematic
phases, and the critical lines between the biaxial and uniaxial nematic
phases, which meet at a Landau multicritical point. For a binary mixture of
biaxial and uniaxial molecules, we have drawn a number of phase diagrams in
terms of temperature versus concentration of biaxial molecules, with fixed
values of $\Delta$, which display many distinct features. Depending on
parameters, there appear additional multicritical points, such as tricritical
and critical end points associated with the biaxial nematic phase. The Landau,
tricritical and critical end points may give rise to a higher-order
multicritical point. Also, depending on the range of values of $\Delta$, there
may be a line of Landau points meeting a line of triple points at a Landau end
point. This intricate behavior of the mixtures of intrinsically uniaxial and
biaxial molecules can be understood in terms of crossings of the microscopic
energy levels as we change the degree of biaxiality $\Delta$.

The explicit expressions for the free energy were used to obtain the
coefficients of an expansion at high temperatures, in the vicinity of the
Landau multicritical point, and to make contact with the Landau-de Gennes
theory. We were then able to check our numerical findings against a number of
phenomenological calculations of the literature. Also, we provided a simple
way of obtaining the expansion coefficients in terms of the values of the
molecular parameters.

The calculations predict the presence of stable biaxial nematic phase at low
temperatures and sufficiently high concentrations of biaxial molecules, which
is in agreement with recent calculations of Longa and coworkers
\cite{LongaPajakWydro} for a more elaborate model system of a mixture of
biaxial molecules in the annealed situation. Depending on the degree of
biaxiality, we predict first-order transitions between biaxial and uniaxial
nematic phases, as well as tricritical points, even in the absence of a Landau
multicritical point. For larger values of the degree of biaxiality, the Landau
point splits into lines of triple points. Also, we note a clear reentrance of
the biaxial regions for some choices of the parameters, which is agreement
with the early work of Alben for a lattice model of platelets \cite{Alben}.

In some recent publications, Akpinar, Reis, and Figueiredo-Neto
\cite{AkpinarEPJE}\cite{AkpinarLC} reported X-ray and optical
characterizations of biaxial nematic structures in a large class of quaternary
liquid-crystalline mixtures. From these measurements, it has been possible to
establish many novel phase diagrams in terms of temperature and molar fraction
of the components, which represents a real advance with respect to the early
work of Yu and Saupe for a ternary lyotropic mixture. For all concentrations
of the amphiphile component, if there is a nematic biaxial structure, it is
thermodynamically stable at intermediate temperatures, between regions of
different uniaxial nematic structures, at lower and higher temperatures. Also
there are examples of temperature-concentration phase diagrams with a clear
indication of the existence of a Landau multicritical point. With a suitable
choice of parameters, the MS6 model of a binary mixtures can qualitatively
reproduce all of these observations. In a very recent experimental
investigation, Amaral and coworkers \cite{Amaral} reanalyzed the phase diagram
of a ternary SDS lyotropic mixture, and pointed out a peculiar coexistence of
uniaxial and biaxial nematic structures, which still seems to demand a
theoretical explanation.

The present calculations, for the elementary MS6 lattice model, at the
mean-field level, are a contribution to the understanding of the effects of
shape fluctuations on the thermodynamic behavior of complex liquid-crystalline
systems. The model of a binary mixture is sufficiently simple to produce a
number of analytical and numerical results for a wide range of values of the
molecular parameters. Also, it seems to be possible to go beyond the
mean-field scenario. The use of more powerful techniques may uncover
additional aspects of the phase diagrams, in particular limitations of the
mean-field approach at very low and very high concentrations.

\acknowledgments

We acknowledge the financial support of the Brazilian agencies FAPESP and CNPq,
as well as of the Brazilian research funding programmes ICNT and NAP on 
Complex Fluids.

\end{document}